\documentclass[preprint,showpacs,superscriptaddress,prd]{revtex4}
\usepackage{graphicx,amsmath,dcolumn,bm}
\begin{document}
\title{\Large{\bf{Systematic analysis of the incoming quark energy loss in cold nuclear matter}}}
\author{Li-Hua Song }
\email[E-mail: ]{songlh@mail.heuu.edu.cn}

\affiliation{Department of Physics, Hebei Normal
             University,
             Shijiazhuang 050024, P.R.China}
\affiliation{Hebei Advanced Thin Films Laboratory, Shijiazhuang
               050024, P.R.China}

\affiliation{ College of Science,  Hebei United University, Tangshan
063009, P.R.China}

\author{Chun-Gui  Duan }
\email[E-mail: ]{duancg@mail.hebtu.edu.cn}

\affiliation{Department of Physics, Hebei Normal
             University,
             Shijiazhuang 050024, P.R.China}
\affiliation{Hebei Advanced Thin Films Laboratory, Shijiazhuang
               050024, P.R.China}


\author{Na  Liu }
\email[E-mail: ]{Liuna@mail.sjzue.edu.cn}

\affiliation{Department of Physics, Hebei Normal
             University,
             Shijiazhuang 050024, P.R.China}
\affiliation{College of Mathematics and Physics, Shijiazhuang
University of Economics, Shijiazhuang 050031, P.R.China}


\begin{abstract}

The investigation into the fast parton energy loss  in cold nuclear
matter is crucial for a good understanding of the parton propagation
in hot-dense medium. By means of four typical sets of nuclear parton
distributions and three  parametrizations of quark energy loss, the
parameter values  in quark energy loss expressions  are determined
from a leading order statistical analysis of the existing
experimental data on nuclear Drell-Yan differential cross section
ratio as a function of the quark momentum fraction. It is found that
with independence on the nuclear modification of parton
distributions, the available experimental data from lower incident
beam energy rule out the incident-parton momentum fraction quark
energy loss. Whether the quark energy loss is linear or quadratic
with the path length is not discriminated. The global fit of all
selected data gives the quark energy loss per unit path length
$\alpha =1.21 \pm 0.09 $ GeV/fm by using nuclear parton distribution
functions determined only by means of the world  data on nuclear
structure function. Our result does not support the theoretical
prediction: the energy loss of an outgoing quark is three times
larger than that of an incoming quark approaching the nuclear
medium. It is desirable that  the present work can provide useful
reference for the Fermilab E906/SeaQuest experiment.


\end{abstract}

\maketitle
\newpage
\vskip 0.5cm

\section*{1  Introduction}

The parton energy loss in high energy collisions has attracted an
increasing amount of  attention from both the nuclear and particle
physics communities for over two decades.   There is a rich
theoretical literature on in-medium parton energy loss extending
back to Bjorken, who proposed the suppressed production of particles
having large transverse momenta, known as jet-quenching, was the
"smoking guns" of  the Quark Gluon Plasma (QGP) formation in high
energy nucleus-nucleus collisions$^{[1]}$.  The wealth of
experimental data on jet-quenching  from RHIC$^{[2,3,4,5,6,7]}$ and
LHC$^{[8]}$ reflect clearly the energy loss of fast partons while
traversing this hot and dense medium. However, a detailed
understanding of the parton energy loss in hot and dense medium
requires  the good investigation into the  fast parton propagation
in cold nuclear matter because  there are  common elements between
the two mediums.

Two sets of experimental data from the semi-inclusive deep inelastic
scattering of lepton on nuclei and the Drell-Yan reaction$^{[9]}$ in
hadron-nucleus collisions can provide the essential information on
the energy loss of fast partons  owing to multiple scattering and
gluon radiation while traversing this cold nuclear medium. The
semi-inclusive deep inelastic scattering on nuclear targets is an
ideal tool to study the energy loss of the outgoing quark in the
cold nuclear medium. In our recent article$^{[10]}$, the
experimental data with quark hadronization occurring outside the
nucleus from HERMES$^{[11]}$ and EMC$^{[12]}$ experiments are picked
out by means of the short hadron formation time. A leading-order
analysis is performed for the hadron multiplicity ratios as a
function of the energy fraction on helium, neon, and copper nuclei
relative to deuteron for the various identified hadrons. It is found
that the theoretical results considering the nuclear modification of
fragmentation functions due to the outgoing quark energy loss are in
good agreement with the selected experimental data.  The obtained
energy loss per unit length is $0.38 \pm 0.03$ GeV/fm for an
outgoing quark by a global fit.

The hadron-induced Drell-Yan reaction on nuclei is an excellent
process to investigate the incoming quark energy loss in cold
nuclear matter because the produced lepton pair does not interact
strongly with the partons in the nucleus. A series of
experiments$^{[13]}$ have been performed at Fermilab and CERN which
presented the Drell-Yan differential cross section distributions  in
order to test the theoretical model, know the momentum distributions
of the projectile and target quarks, and explore the nuclear target
dependence. Four experimental collaborations have measured the
Drell-Yan differential cross section ratio of two different nuclear
targets bombarded by the same hadron at the same centre-of-mass
energy in order to study the nuclear effects on Drell-Yan reaction.
They are NA3$^{[14]}$ and NA10$^{[15]}$
 Collaborations from CERN, and E772$^{[16]}$  and
E866$^{[17]}$ Collaborations from Fermilab. The advantage of using
the Drell-Yan differential cross section ratio at the same energy is
that the differential cross section ratio can reduce the dependence
on the beam hadrons, and cancel the most uncertainties regarding the
lepton pair production. Additionally, the differential cross section
ratio can avoid the influence of the QCD next-to-leading order
correction. Theoretically, it has proved that the effect of
next-to-leading order correction on the Drell-Yan differential cross
section ratio as a function of the quark momentum fraction  can be
negligible for the 800 GeV proton beam at Fermilab and  lower energy
beam$^{[18]}$.

In our previous articles$^{[19,20]}$,  the energy loss effect  on
the Fermilab E866 nuclear Drell-Yan differential cross section ratio
was investigated  as a function of the quark momentum fraction of
the beam proton at the hadron level in the framework of the Glauber
model, and at parton level by using two typical kinds of quark
energy loss parametrization, respectively. It was confirmed that the
energy loss effect can suppress evidently the differential cross
sections versus the quark momentum fraction.  In the recent
work$^{[21]}$, the study on quark energy loss is extended to the
E772 data without performing the global fit to E772 and E866 data.
It is found that the quark energy loss effect on nuclear Drell-Yan
cross section ratio becomes greater with the increase of quark
momentum fraction in the target nuclei. The global analysis of
nuclear parton distribution functions including E772 data
overestimates the nuclear modification in the sea quark distribution
if the quark energy loss effect is neglected. It is noticeable that
the E772 and E866 Collaborations used the same 800 GeV proton beam
incident on various nuclei. The measured momentum fraction of the
target parton is in the range $0. 01 < x_2 \leq 0.271$.

The main goal of the present work is to extract the incoming quark
energy loss in cold nuclear matter systematically from a global
analysis of these experimental results on nuclear Drell-Yan
differential cross section ratio from NA3$^{[14]}$ and NA10$^{[15]}$
Collaborations at CERN, and E772$^{[16]}$  and E866$^{[17]}$
Collaborations at Fermilab. The main improvements over our earlier
work are twofold: on one hand, the  error estimate for the incoming
quark energy loss is presented, and on the other hand, by adding the
NA3$^{[14]}$ and NA10$^{[15]}$ data from CERN, the used experimental
data can cover 140 GeV, 150 GeV, 286 GeV and 800 GeV incident hadron
beam with the target-quark momentum fraction from 0.01 to 0.45. The
extended beam energy and kinematic coverage significantly increase
the sensitivity to incident parton energy loss and nuclear
modification in the sea quark distribution.  It is hoped to provide
a good understanding of the parton energy loss in cold nuclear
matter from the available data, and to facilitate the theoretical
research on the energy loss of an incoming quark and outgoing quark
in nuclear matter.

This article is organized as follows. A brief formalism for  the
differential cross section in nuclear Drell-Yan process is detailed
in Sect.2, followed by the data selection  in Sect.3. The obtained
results are discussed in Sect.4. Finally, the summary and concluding
remarks are given in Sect.5.

\section*{2  Dilepton production differential cross section in nuclear targets }


At the leading order(LO) in perturbation theory, the lepton pair
production differential cross section in hadron-nucleus collisions
can be obtained from the convolution of differential partonic cross
section $\bar{q}q \rightarrow l^{+}l^{-}$ with  the parton
distribution functions in the incident hadron $h$ and the target
nucleus $A$. With neglecting the incoming quark energy loss in cold
nuclear matter, the differential cross section is written as
\begin{equation}
 \frac{d^2\sigma}{dx_1dx_2}=\frac{4\pi\alpha_{em}^2}{9sx_1x_2}
 \sum_{f}e^2_f[q^h_f(x_1,Q^2)\bar{q}^A_f(x_2,Q^2)
 +\bar{q}^h_f(x_1,Q^2)q^A_f(x_2,Q^2)],
\end{equation}
where $x_1$($x_2$) is the momentum fraction of the partons in the
beam hadron(target),  $ \alpha _{em}$ is the fine structure
constant, $\sqrt{s}$ is the center of mass energy of the hadronic
collision, $e_f$ is the charge of the quark with flavor $f$, $Q^2$
is the invariant mass of a lepton pair,  $q^{h(A)}_{f}(x,Q^2)$ and
${\bar q}^{h(A)}_{f}(x,Q^2)$ are respectively the quark and
anti-quark distribution function with Bjorken variable $x$ and
photon virtuality $Q^2$ in the hadron (nucleon in the nucleus A),
and the sum is carried out over the light flavor.

In the hadron-induced Drell-Yan reaction on nuclei, the incoming
quark can lose its energy $\Delta E_{q}$, owing to multiple
scattering on the surrounding nucleon and gluon radiation while
propagating through the nucleus. The energy loss  of an incoming
quark results in an average change in its momentum fraction prior to
the collision, $ \Delta x_1= \Delta E_{q}/E_{h}$, where $E_{h}$ is
the incident hadron energy. On the basis of theoretical research,
three  parametrizations for quark energy loss have been proposed
separately by Brodsky and Hoyer$^{[22]}$, Baier et al.$^{[23]}$, and
by  Gavin and Milana$^{[24]}$. One is $\Delta x_1= {\alpha}
<L>_A/E_h$, where $\alpha$ denotes the incident quark energy loss
per unit length in nuclear matter, $<L>_A=3/4(1.2A^{1/3})$fm is the
average path length of the incident quark in the nucleus A.  Another
one is  $\Delta x_1= {\beta}<L>^2_A/E_h$. Obviously, the quark
energy loss is quadratic with the path length. In what follows, the
two different parametrizations  are called the linear and quadratic
quark energy loss,  respectively. The third form is $\Delta x_1=
{\kappa}x_1A^{1/3}$, which is  named  the incident-parton momentum
fraction quark energy loss. In these three expressions, $\alpha,
\beta$ and $\kappa$ can be extracted by a global analysis to nuclear
Drell-Yan experimental data on the differential cross section ratio,
respectively.

The quark energy loss in target nucleus shifts the incident quark
momentum fraction from $x'_1=x_1+\Delta x_1$ to $x_1$ at the point
of fusion. With adding the  quark energy loss in the nucleus, the
nuclear Drell-Yan differential cross section can be expressed as
\begin{equation}
 \frac{d^2\sigma}{dx_1dx_2}=\frac{4\pi\alpha_{em}^2}{9sx_1x_2}
 \sum_{f}e^2_f[q^h_f(x'_1,Q^2)\bar{q}^A_f(x_2,Q^2)
 +\bar{q}^h_f(x'_1,Q^2)q^A_f(x_2,Q^2)].
\end{equation}


The parton distribution functions inside a nucleus from the
 Drell-Yan differential cross section have been found to differ
notably from the corresponding ones in the free nucleon with the
discovery of the nuclear EMC effect some twenty years ago(see
Ref.[25], and references therein). Despite a significant worldwide
effort in experiment and theory, there is as yet no consensus
concerning the origin of this effect. In view of the importance for
finding any new physical phenomena in the high-energy nuclear
reactions, the global analyses of nuclear parton distribution
functions, which parallel those for the free proton, have been
performed in the past decade by different groups:
HKM/HKN07$^{[26,27]}$, nDS$^{[28]}$, and EPS09$^{[29]}$.  The four
sets of nuclear parton distribution functions employed the existing
experimental data on nuclear structure functions  from the electron
and muon deep inelastic scattering. Unfortunately, the nuclear
structure functions are composed of nuclear sea and valence quark
distributions. The fact results in that the nuclear valence quark
distributions are relatively well determined except for the small
Bjorken variable $x$  region, and nuclear antiquark distributions
 in small  $x$ region. It is difficult to
constraint the antiquark distributions at medium and large $x$
region. It is expected that the nuclear Drell-Yan experimental data
can pin down the nuclear valence quark distributions in the small
$x$  region and nuclear antiquark distributions in the medium $x$
region $0.01 <  x <  0.3$. For this reason, HKN07 and EPS09 added
Fermilab E772 and E866 nuclear Drell-Yan data,  and nDS included
E772 experimental data with difference from HKM.

\begin{figure}[!]
\centering
\includegraphics*[width=11cm, height=10cm]{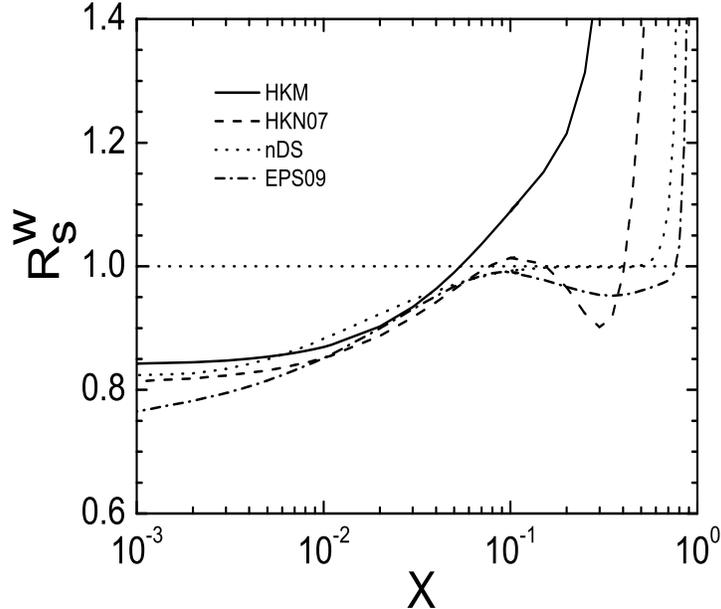}
\vspace{-0.5cm} \caption{The nuclear  modification of sea quark
distribution at $Q^2=50 GeV^2$ as a function of Bjorken variable $x$
for tungsten nucleus. The solid, dashed, dotted and dash dot lines
correspond to the the results given by  HKM, HKN07, nDS and EPS09
nuclear parton distributions, respectively.}
\end{figure}

The quantitative comparison between the different sets of nuclear
parton distribution functions shows that the nuclear  modification
for valence quarks  agrees nicely in the large-$x$ region $x >0.3$.
In other $x$ region, HKN07, nDS and EPS09 give nuclear modifications
relatively close to each other. Only the HKM displays a smaller
antishadowing in the region $0.01<x < 0.3$ than other sets, and no
nuclear correction in $x < 0.01$. The nuclear modification of sea
quark distribution for tungsten nucleus, $R_s^{\rm W}(x, Q^2=50
GeV^2)$ is shown in Fig.1, as a function of Bjorken variable $x$
from the leading order HKM (solid line), HKN07 (dashed line), nDS
(dotted line), and EPS09 (dash dot line) nuclear effects. It is
found that the nuclear modifications from different sets are
relatively close to each other in the  region $0.01 < x < 0.08$,
however gives a clear deviation in the  region $x>0.08$. It is
apparent that the nuclear modifications with including nuclear
Drell-Yan data do not give a good consistency in the medium $x$
region from HKN07, nDS and EPS09 parameterizations. The fact goes
contrary to one's wishes.

\section*{3  The experimental data}

 In  our present analysis, the
experimental data, providing the input for the value of the
parameter in the three representative expressions of quark energy
loss,  are taken from NA3$^{[14]}$ and NA10$^{[15]}$ Collaborations
at CERN, and E772$^{[16]}$  and E866$^{[17]}$ Collaborations at
Fermilab. The experimental data sets used here are summarised in
Table I, in which the beam energy $E_{beam}$ of incident hadron, the
projectile/target species, the covered domain on the momentum
fraction of the hadron and target parton,  and the number N of
points in each data sample are specified.  In total, our analysis
has 269 data points, and 6 nuclei from Beryllium up to Platinum.

\begin{table}[htb]
  \begin{center}
   \caption{Experimental data sets selected for the present analysis.}
  \begin{tabular}[c]{p{1.0cm}ccccccc}
    \hline
    \hline
    Exp. & $E_{beam}$ (GeV) & Proj.  & Target & $x_1$& No. data  & $x_2$& No. data   \\
    \hline
    NA10a$^{[15]}$ & 140 & $\pi^-$ & D, W       &0.39--0.82  &5  & 0.163--0.360 & 4\\
    NA3$^{[14]}$ & 150  & $\pi^-$ & H, Pt      &0.25--0.95  & 8 & 0.074--0.366& 7\\
    NA10b$^{[15]}$ & 286 & $\pi^-$ & D, W       & 0.22--0.83 &  9 & 0.125--0.451& 6\\
    E772$^{[16]}$ & 800 & $p$ & D, C, Ca, Fe, W& 0.15--0.85 &122  & 0.04--0.271 & 36\\
    E866$^{[17]}$ & 800 & $p$ & Be, Fe, W    & 0.21--0.95 & 56 & 0.01--0.12& 16 \\
    \hline
    \hline
\end{tabular}
  \end{center}
\end{table}

To be emphasized,   NA3 and NA10 data used cover the momentum
fraction of the target parton from 0.074 to 0.366, and from 0.125 to
0.451, respectively. In the region $x_2 > 0.1$, the nuclear
modification to sea quark distribution given by the different sets
displays a gradually large deviation from each other with the
increase of  the momentum fraction of the target parton. Therefore,
NA3 and NA10 data can show better the difference between  HKM,
HKN07, nDS and EPS09 parameterizations. Meanwhile, the bigger range
coverage on  beam energy and  momentum fraction of the target parton
can make us to find very clearly the quark energy loss effect on the
nuclear Drell-Yan differential cross section ratio.

\section*{4 Results and discussion }

In order to study the quark energy loss effect in the hadron-induced
Drell-Yan reaction on nuclei,  determine the values of the
parameters $\alpha$, $\beta$ and $\kappa$ in quark energy loss
expressions, and investigate the dependence of quark energy loss on
the nuclear parton distribution functions, we calculate in
perturbative QCD  leading order(LO)  the Drell-Yan cross section
ratio $R^{theo}_{A_{1}/A_{2}}$ on two different nuclear targets
bombarded by hadron
\begin{equation}
R^{theo}_{A_{1}/A_{2}}(x_{1(2)})=\int
dx_{2(1)}\frac{d^2\sigma^{h-A_{1}}}{dx_1dx_2} /\int
dx_{2(1)}\frac{d^2\sigma^{h-A_{2}}}{dx_1dx_2}.
\end{equation}
The comparison is performed   with selected experimental data on the
Drell-Yan differential cross section ratio. The integral range in
above equation is obtained by means of the relative experimental
kinematic region with neglecting the nuclear modifications in
Deuterium. In our calculation as following, we use the four sets of
leading order nuclear parton distribution functions together with
CTEQ6L parton density in the proton$^{[30]}$, and parton density in
the negative pion$^{[31]}$.


\begin{table}[t,m,b]
\caption{The $\chi^2$/N-values computed using HKM, HKN07, nDS and
EPS09  nuclear parton distribution functions without quark energy
loss effect. The notation $x_1$ and $x_2$ indicate the momentum
fraction of the incident hadron and target parton,respectively. }
\begin{ruledtabular}
\begin{tabular*}{\hsize}
{c@{\extracolsep{0ptplus1fil}} c@{\extracolsep{0ptplus1fil}}
c@{\extracolsep{0ptplus1fil}} c@{\extracolsep{0ptplus1fil}}
c@{\extracolsep{0ptplus1fil}}}
    Exp. data & HKM  & HKN07& nDS & EPS09  \\
    \hline
     NA10a($x_1$) & 23.69 & 6.84 & 8.81 &5.03  \\
    NA10a($x_2$) & 25.68 & 5.26 & 10.42 & 4.92 \\
    NA3($x_1$) & 5.99  & 4.81 & 4.45  &3.68  \\
    NA3($x_2$) & 9.93 &6.98 & 7.67 &6.35 \\
    NA10b($x_1$) & 3.81 & 1.45 & 1.60 & 1.42 \\
    NA10b($x_2$) & 5.11 & 0.65 & 1.90 & 0.86 \\
     E772($x_1$) & 1.92 &1.41 &1.47& 1.33  \\
    E772($x_2$)& 4.83 & 1.58 & 1.79& 0.82 \\
    E866($x_1$) & 1.44& 0.90 & 1.19 & 0.84  \\
    E866($x_2$) & 3.00 &1.17 & 2.17  & 0.95  \\
\end{tabular*}
\end{ruledtabular}
\end{table}

When  the incoming quark energy loss effect is neglected in the
hadron-induced Drell-Yan reaction on nuclei, the calculated results
are compared with the experimental data  selected for our analysis
on the Drell-Yan differential cross section ratio as a function of
the momentum fraction of the incident hadron and target parton. The
$\chi^2$/N (N being the number of data points) computed are
summarized in Table II by means of  HKM, HKN07, nDS and EPS09
nuclear parton distribution functions, respectively. Usually, if the
$\chi^2$/N  is not much larger than one, the theoretical results are
considered as being statistically consistent with the experimental
data. As seen from the Table II,  each analysis on nuclear effects
has consistently much larger $\chi^2$/N values for NA3 and NA10a
data from the negative pion incident Drell-Yan reaction on nuclei.
Interestingly, very larger $\chi^2$/N value is
 23.69 and 25.68 for  NA10a($x_1$) and NA10a($x_2$) data,
 respectively, from the HKM nuclear effects.


\begin{table}[t,m,b]
\caption{The values of $\alpha$, $\beta$ and $\kappa$, $\chi^2/ndf$
and S factors extracted from each data sample with  HKM nuclear
corrections. The bottom row corresponds to the global fit of all
selected data. }
\begin{ruledtabular}
\begin{tabular*}{\hsize}
{c@{\extracolsep{0ptplus1fil}} c@{\extracolsep{0ptplus1fil}}
c@{\extracolsep{0ptplus1fil}} c@{\extracolsep{0ptplus1fil}} }
    Exp. data & $\alpha$ ( $\chi^{2}/ndf$, \hspace{0.3mm}S)  &
    $\beta$ ( $\chi^{2}/ndf$, \hspace{0.3mm}S)& $\kappa$ ( $\chi^{2}/ndf$, \hspace{0.3mm}S) \\
    \hline
    NA10a($x_1$) & $1.53 \pm0.14$(1.61,1.27)& $0.248 \pm0.022$(1.49,1.22)&$0.0152\pm 0.0015$(4.12,2.03)   \\
    NA10a($x_2$) &  $1.46 \pm0.16$(0.81,1.00)& $0.230\pm 0.025$(0.74,1.00)&$0.0210 \pm0.0030$(2.69,1.64)   \\
    NA3($x_1$) & $1.82 \pm0.29$(1.69,1.30)& $0.310 \pm0.050$(1.44,1.20)&$0.0120 \pm0.0040$(3.40,1.84)    \\
    NA3($x_2$) &  $1.85\pm 0.36$(2.53,1.59)& $0.300 \pm0.060$(2.24,1.50)&$0.0130\pm 0.0050$(5.78,2.40)   \\
    NA10b($x_1$) &  $0.88 \pm0.20$(1.31,1.14)& $0.140 \pm0.030$(1.31,1.14)&$0.0070\pm 0.0015$(1.47,1.21)   \\
    NA10b($x_2$) &  $0.79 \pm0.14$(0.45,1.00)& $0.125 \pm0.023$(0.45,1.00)&$0.0083 \pm0.0016$(0.88,1.00)  \\
    E772($x_1$) &$1.26\pm 0.16$(1.33,1.15)& $0.230 \pm0.030$(1.35,1.16)&$0.0042 \pm0.0005$(1.26,1.12)   \\
    E772($x_2$)&  $1.29 \pm0.13$(1.10,1.05)& $0.230 \pm0.030$(1.53,1.24)&$0.0066\pm 0.0006$(1.69,1.30)   \\
    E866($x_1$) & $1.28 \pm0.22$(0.79,1.00)& $0.190\pm 0.030$(0.80,1.00)&$0.0026 \pm0.0004$(0.77,1.00)   \\
    E866($x_2$) & $1.27 \pm0.23$(1.12,1.06)& $0.190\pm 0.030$(1.13,1.06)&$0.0035 \pm0.0006$(0.78,1.00)   \\
    Global fit &  $1.21 \pm0.09$(1.07,1.03)& $0.190\pm 0.020$(1.11,1.05)&$0.0037 \pm0.0003$(1.21,1.10)   \\
\end{tabular*}
\end{ruledtabular}
\end{table}

To determinate  the optimal parameter from each experimental data
set, we adopt the $\chi^2$ analysis method described in Ref.[10].
The obtained results  are summarized in Table III. by combining the
HKM cubic type of nuclear parton distributions with linear,
quadratic and incident-parton momentum fraction quark energy loss,
respectively. The values of $\alpha$, $\beta$, $\kappa$ extracted
from the individual fit of each data sample, as well as their
corresponding rescaled  error, $\chi^2$ per number of degrees of
freedom ($\chi^2/ndf$) and S factors, are listed in Table III, in
which  the bottom row corresponds to the global fit of all selected
data. As can be found from Table III compared with Table II,
$\chi^2$ values with quark energy loss effect, overall,  are much
smaller than those with only HKM nuclear effects of parton
distributions. The agreement of theoretical calculations with
experimental data has a significant improvement. However, the
computed results with the incident-parton momentum fraction quark
energy loss have yet a significant deviation from NA3 and NA10a data
sets.

Regarding the linear, quadratic  and the incident-parton momentum
fraction quark energy loss, the global fit of all data makes $\alpha
=1.21 \pm 0.09 $ with the relative uncertainty $\delta \alpha /
\alpha \simeq 7\%$ and $\chi^2/ndf = 1.07$, $\beta =0.19 \pm 0.02 $
with  $\delta \beta / \beta \simeq 10\%$ and $\chi^2/ndf = 1.11$,
and $\kappa =0.0037 \pm 0.0003 $ with  $\delta \kappa / \kappa
\simeq 8\%$ and $\chi^2/ndf = 1.21$.

\begin{figure}[!]
        \includegraphics*[width=70mm]{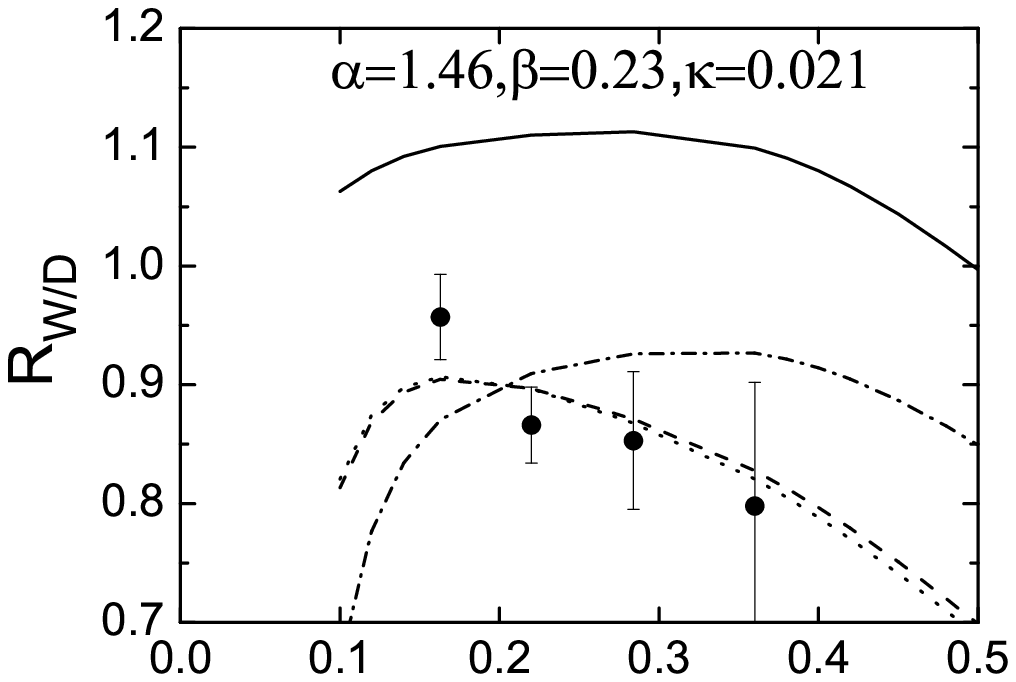} \hspace{0.5mm}
        \includegraphics*[width=70mm]{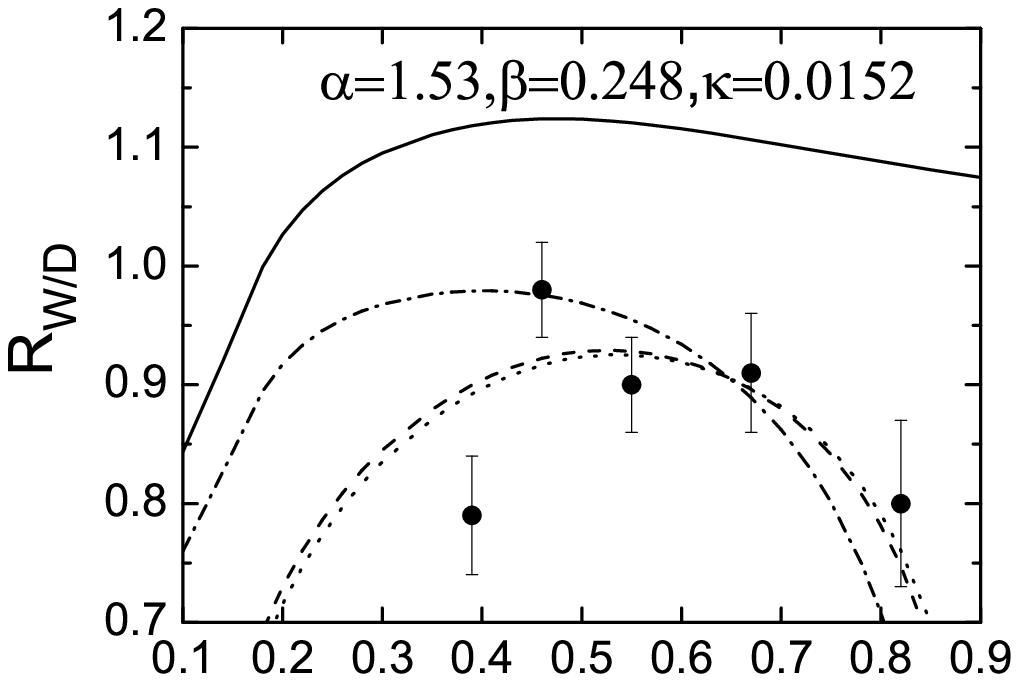} \\
\vspace{-1.0cm}
        \includegraphics*[width=70mm]{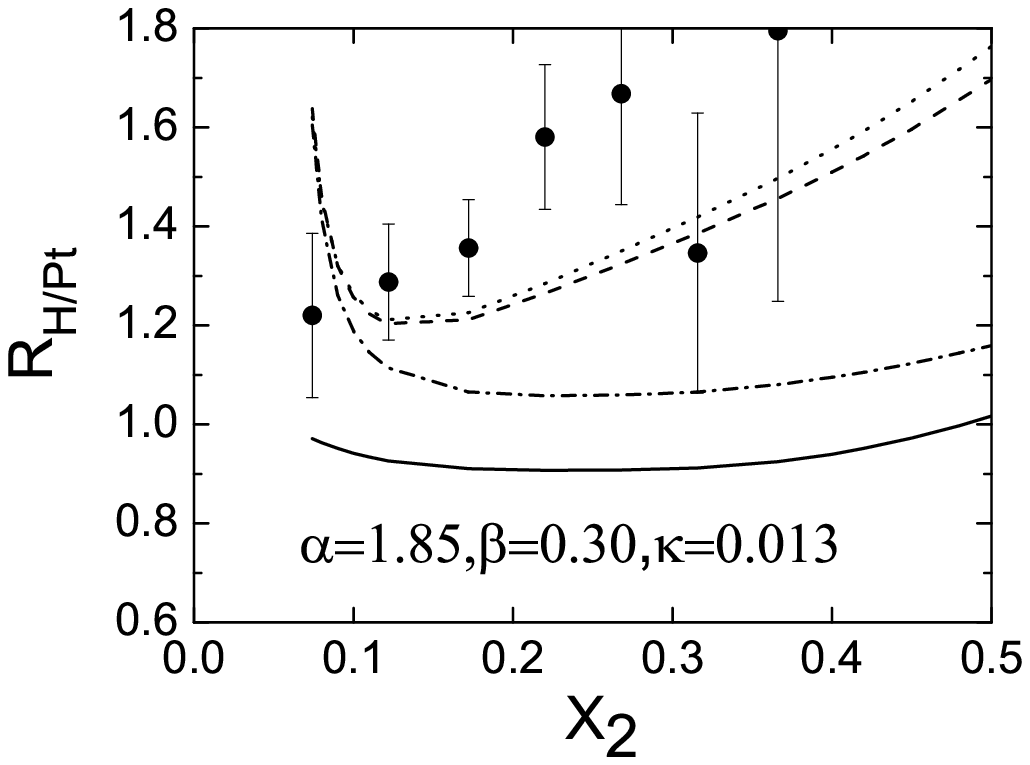} \hspace{0.5mm}
        \includegraphics*[width=70mm]{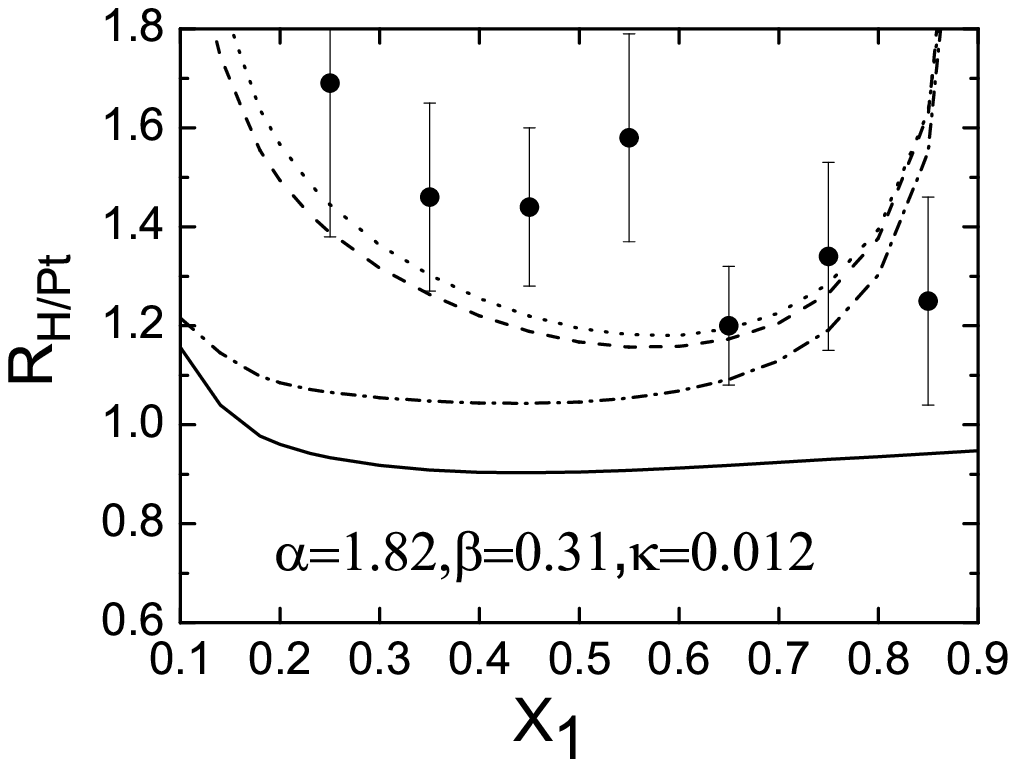}
       \vspace{-0.25cm}
 \caption{ The nuclear Drell-Yan cross section ratios
$R_{A_{1}/A_{2}}(x_{1})$ and $R_{A_{1}/A_{2}}(x_{2})$ by using HKM
nuclear parton distributions. The solid curves correspond to the
results with only  nuclear effects of parton distributions. The
dashed, dotted and dash dot curves show the combination of nuclear
effects of parton distributions with linear, quadratic and
incident-parton momentum fraction quark energy loss, respectively.
The relative optimal parameter is taken from the fit to
corresponding data sample. The experimental data are taken from
NA3$^{[14]}$ and NA10a$^{[15]}$.}
\end{figure}

To demonstrate intuitively the energy loss effect of an incoming
quark on the nuclear Drell-Yan cross section ratio, the calculated
results combining HKM cubic type of nuclear parton distributions are
compared with NA10a and NA3 data in Fig.2. It is necessary to note
that NA3 Collaboration provided the negative pion incident Drell-Yan
cross section ratio $R_{H/Pt}$ as a function of parton momentum
fraction. If the ratio $R_{H/Pt}$ is transformed to $R_{Pt/H}$, the
ratio $R_{Pt/H}$ has the same tendency as $R_{W/D}$ at the beam
energy 140 GeV.  It can be seen from the Fig.2 that the calculated
differential cross section ratios are nearly same from the linear
and quadratic quark energy loss.  The theoretical prediction from
the incident-parton momentum fraction quark energy loss is not in
agreement with the NA3, NA10a$(x_2)$  and NA10a$(x_1)$ data in
$x_1<0.4$. Therefore,  we can conclude apparently that the existing
experimental data from lower incident beam energy  rule out the
possibility of the incident-parton momentum fraction quark energy
loss. Whether the quark energy loss is linear or quadratic with the
path length is  not determined.


\begin{table}[t,m,b]
\caption{The values of $\alpha$, $\beta$ and $\kappa$ with
$\chi^2/ndf$ and S factors extracted from the global fit of all data
by using HKN07, nDS and EPS09 nuclear corrections, respectively.  }
\begin{ruledtabular}
\begin{tabular*}{\hsize}
{c@{\extracolsep{0ptplus1fil}} c@{\extracolsep{0ptplus1fil}}
c@{\extracolsep{0ptplus1fil}} c@{\extracolsep{0ptplus1fil}} }
    &$\alpha^{HKN07}$ ( $\chi^{2}/ndf$, \hspace{0.3mm}S)
    & $\alpha^{nDS}$ ( $\chi^{2}/ndf$, \hspace{0.3mm}S)& $\alpha^{EPS09}$ ( $\chi^{2}/ndf$, \hspace{0.3mm}S) \\
    \hline
    Global fit & $ 0.64\pm0.09$(1.09,1.04)& $ 0.73 \pm0.09$(1.08,1.04)&$ 0.23 \pm0.07$(1.05,1.02)   \\
  \hline\hline
    & $\beta^{HKN07}$ ( $\chi^{2}/ndf$, \hspace{0.3mm}S)  &
    $\beta^{nDS}$ ( $\chi^{2}/ndf$, \hspace{0.3mm}S) & $\beta^{EPS09}$ ( $\chi^{2}/ndf$, \hspace{0.3mm}S) \\
 \hline
 Global fit & $ 0.103\pm0.014$(1.09,1.04)& $ 0.122 \pm0.013$(1.08,1.04)&$ 0.042 \pm0.015$(1.05,1.02)   \\
  \hline\hline
     & $\kappa^{HKN07}$ ( $\chi^{2}/ndf$, \hspace{0.3mm}S)  &
  $\kappa^{nDS}$ ( $\chi^{2}/ndf$, \hspace{0.3mm}S)& $\kappa^{EPS09}$ ( $\chi^{2}/ndf$, \hspace{0.3mm}S) \\
    \hline
Global fit & $ 0.0023\pm0.0003$(1.08,1.04)& $ 0.0026 \pm0.0004$(1.08,1.04)&$ 0.0009 \pm0.0004$(1.05,1.02)   \\
 \end{tabular*}
\end{ruledtabular}
\end{table}

\begin{figure}[!]
        \includegraphics*[width=70mm]{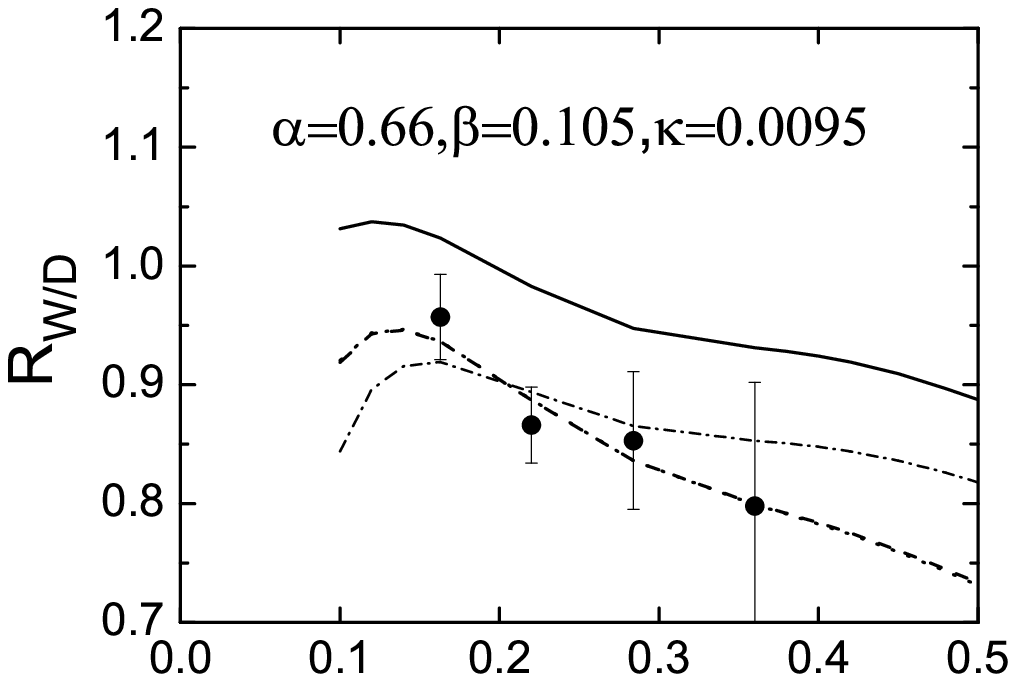} \hspace{0.5mm}
        \includegraphics*[width=70mm]{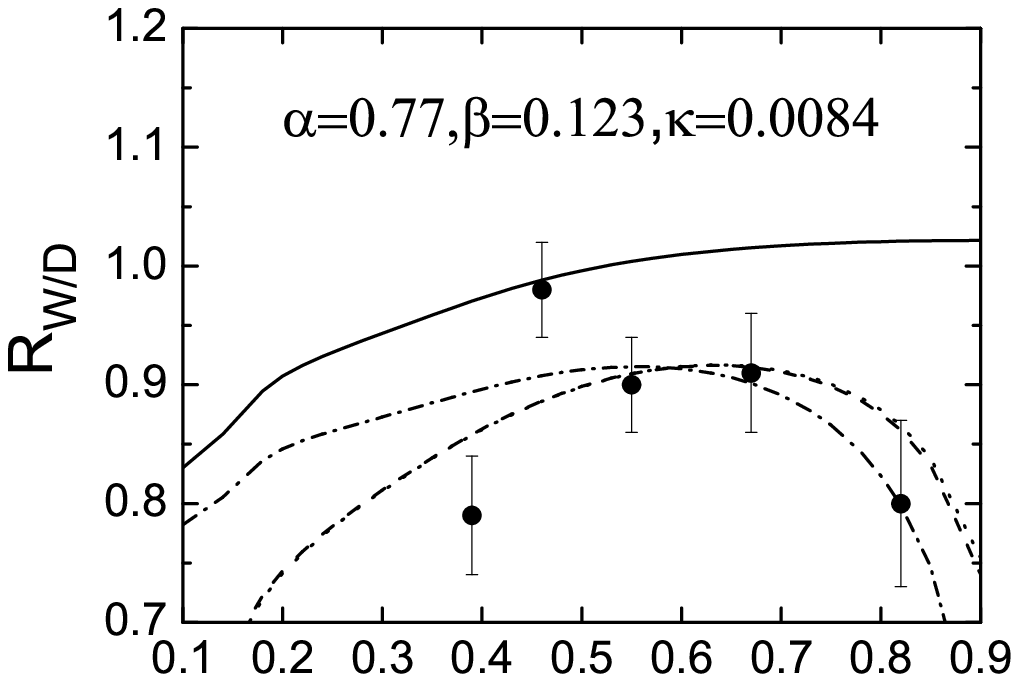} \\
\vspace{-1.0cm}
        \includegraphics*[width=70mm]{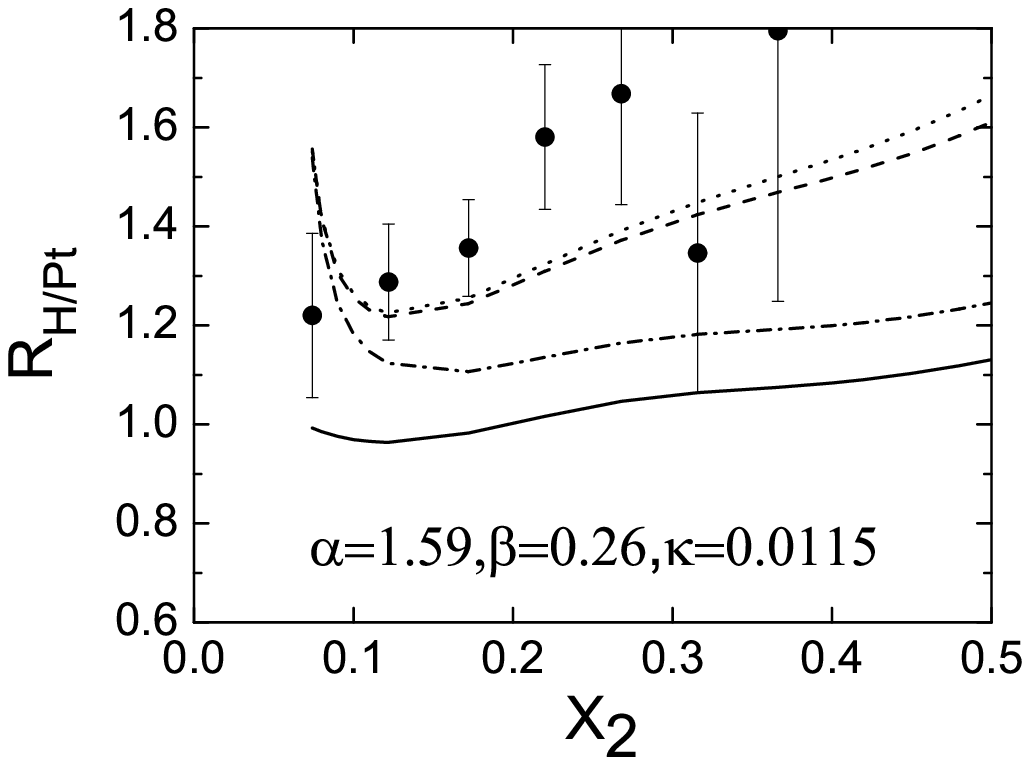} \hspace{0.5mm}
        \includegraphics*[width=70mm]{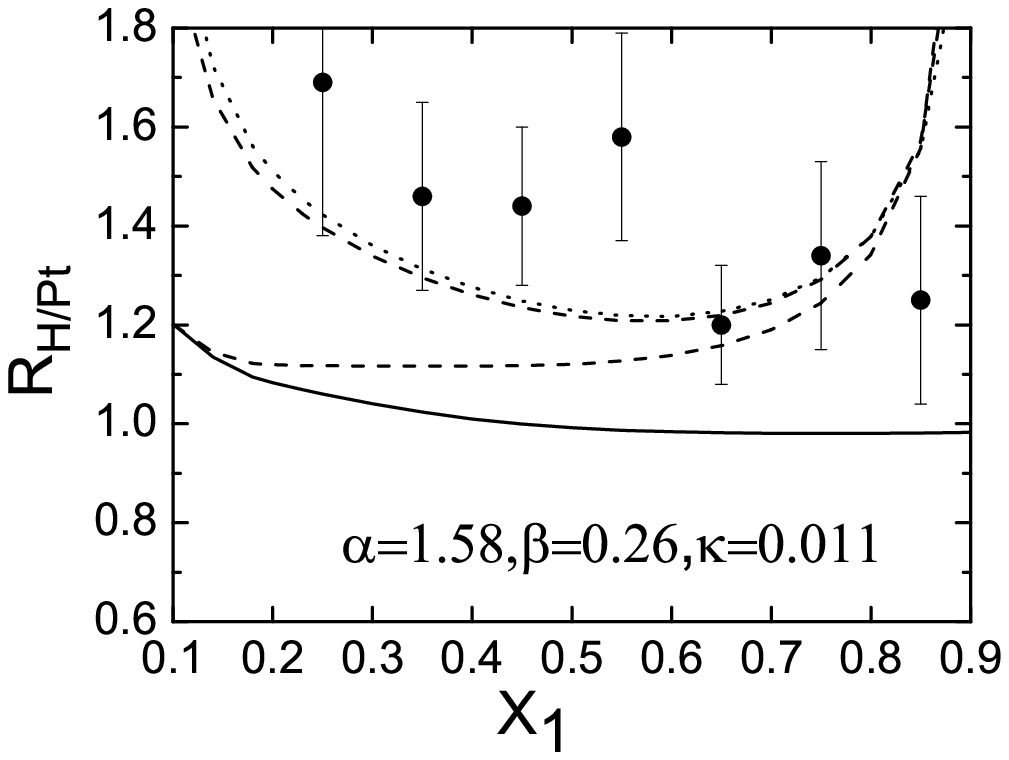} \\
\vspace{-1.0cm} \caption{ The nuclear Drell-Yan cross section ratios
$R_{A_{1}/A_{2}}(x_{1})$ and $R_{A_{1}/A_{2}}(x_{2})$ by using HKN07
nuclear parton distributions. The other comments are the same as
those in Fig.2.}
\end{figure}

\begin{figure}[t,m,b]
        \includegraphics*[width=70mm]{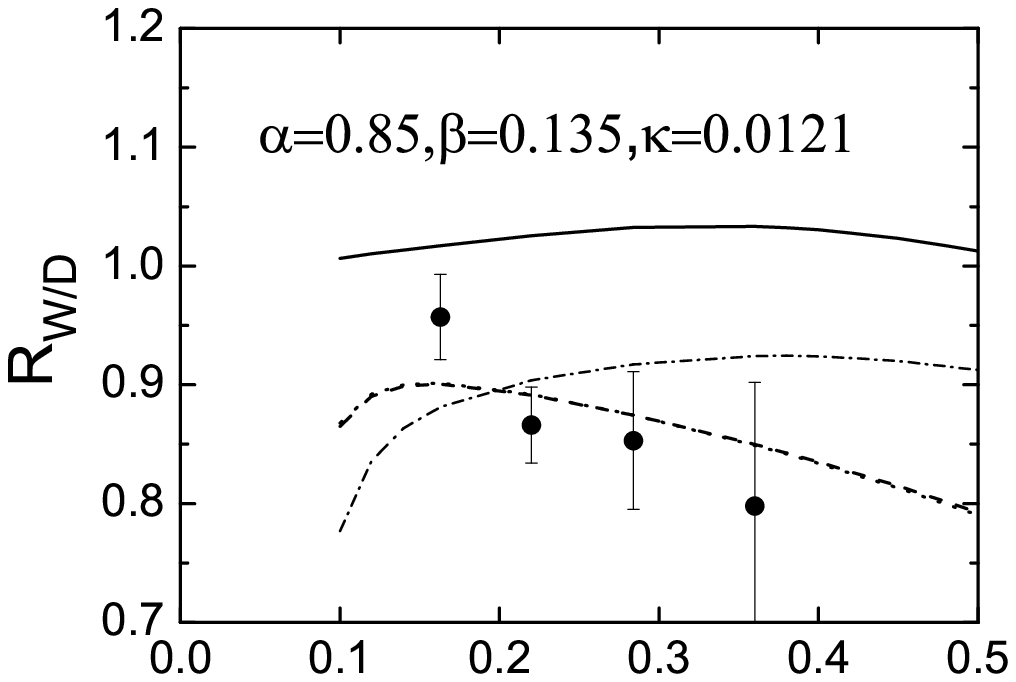} \hspace{0.5mm}
        \includegraphics*[width=70mm]{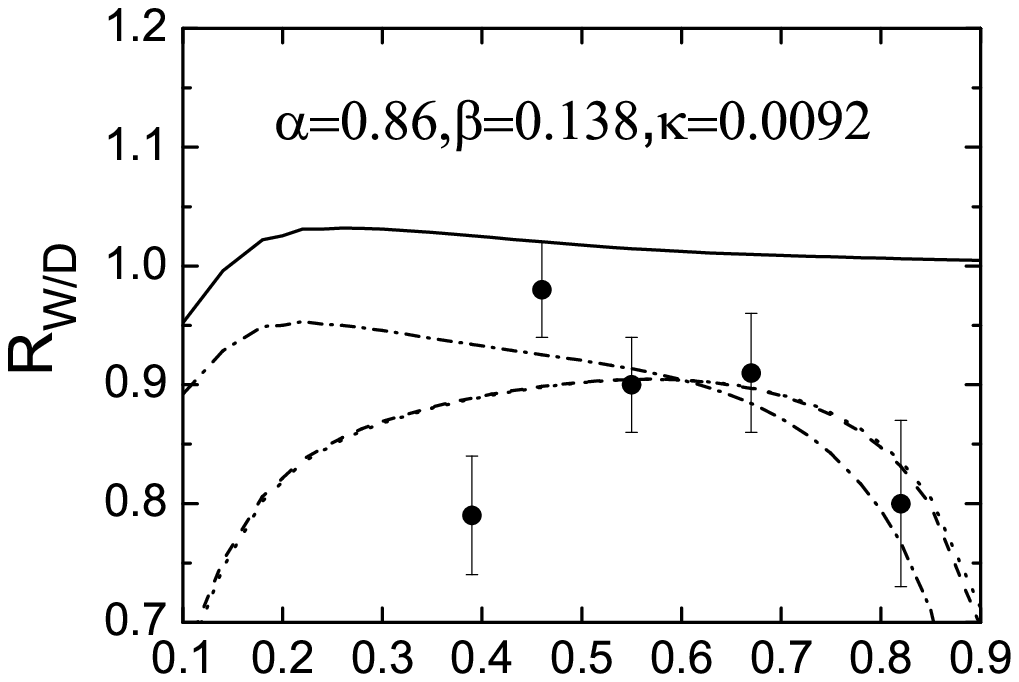} \\
\vspace{-1.0cm}
        \includegraphics*[width=70mm]{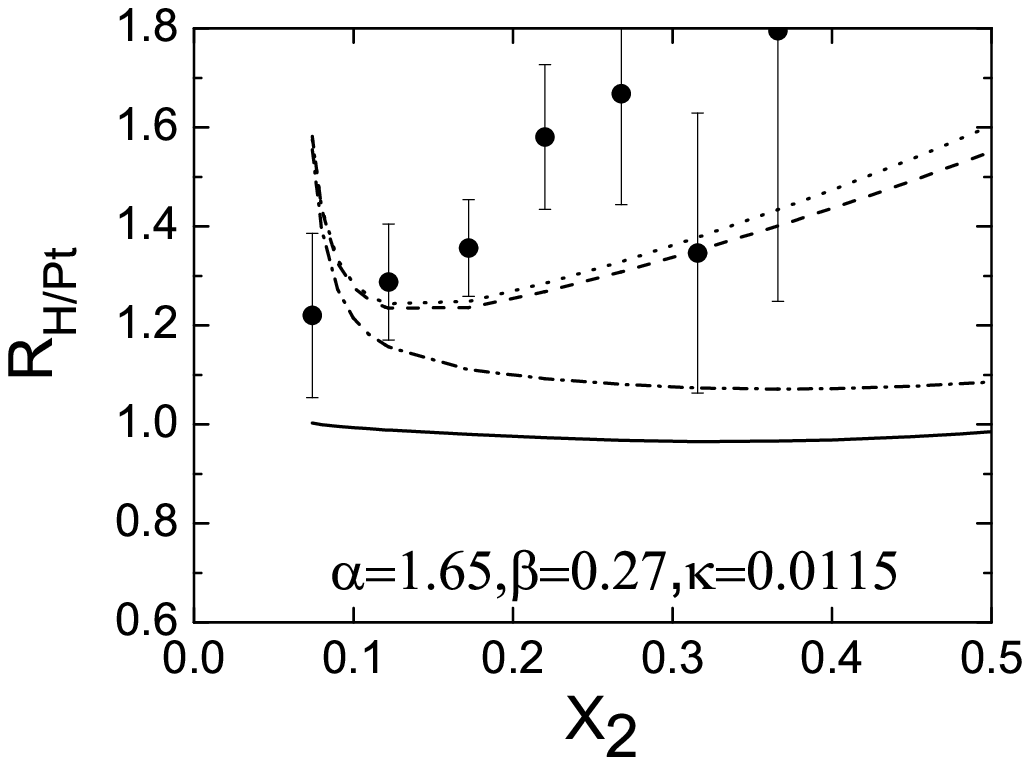} \hspace{0.5mm}
        \includegraphics*[width=70mm]{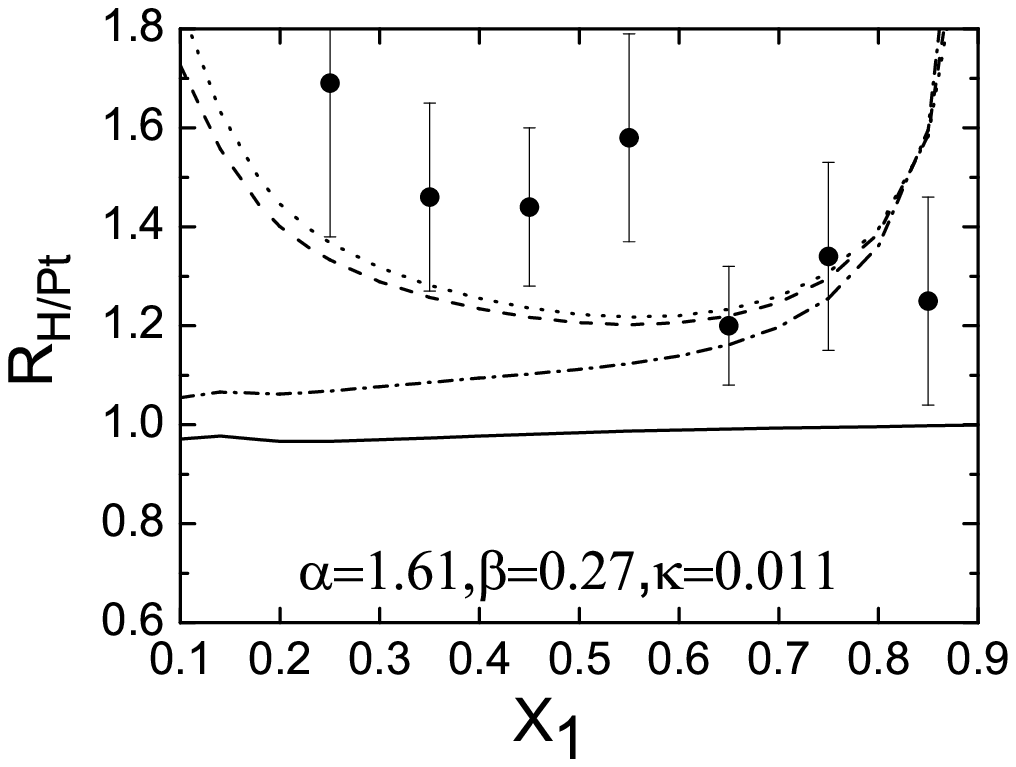} \\
\vspace{-1.0cm} \caption{ The nuclear Drell-Yan cross section ratios
$R_{A_{1}/A_{2}}(x_{1})$ and $R_{A_{1}/A_{2}}(x_{2})$ by using nDS
nuclear parton distributions.  The other comments are the same as
those in Fig.2.}
\end{figure}

\begin{figure}[t,m,b]
        \includegraphics*[width=70mm]{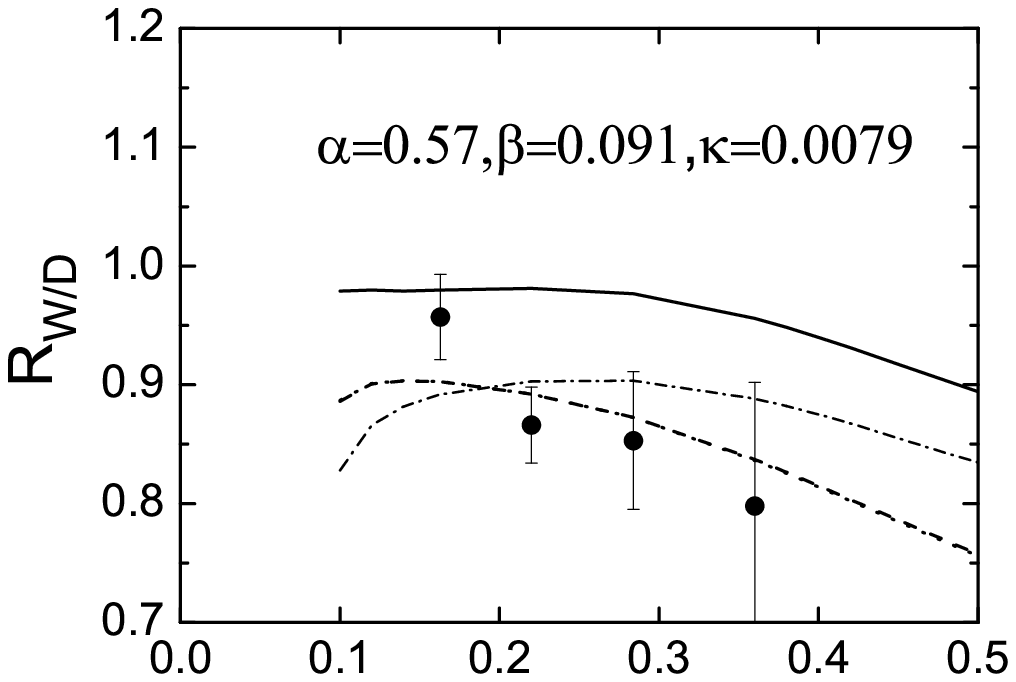} \hspace{0.5mm}
        \includegraphics*[width=70mm]{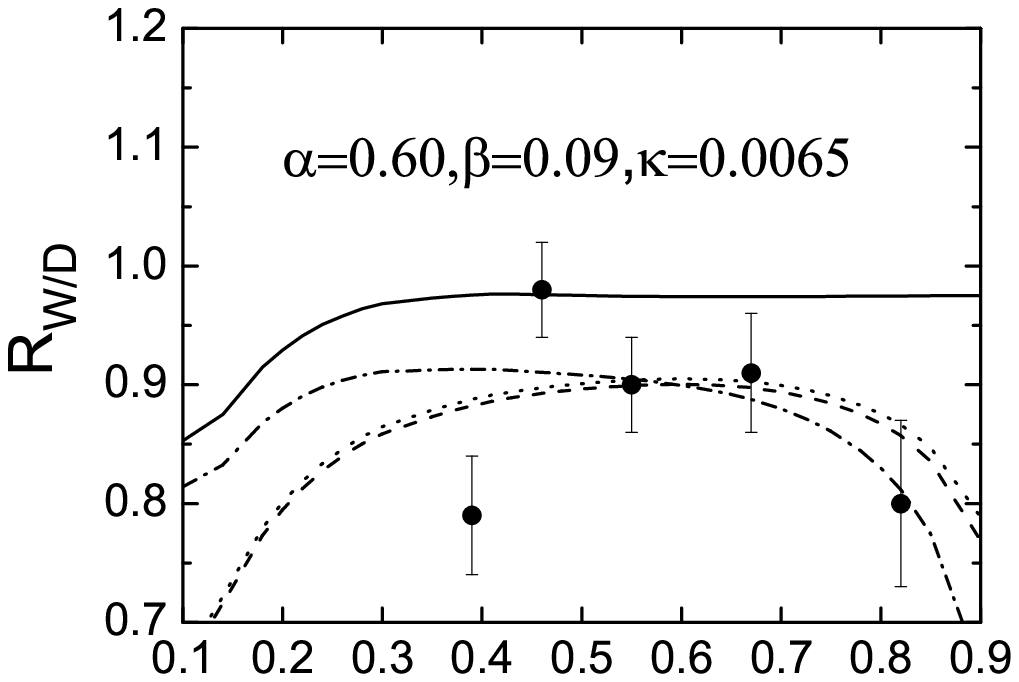} \\
\vspace{-1.0cm}
        \includegraphics*[width=70mm]{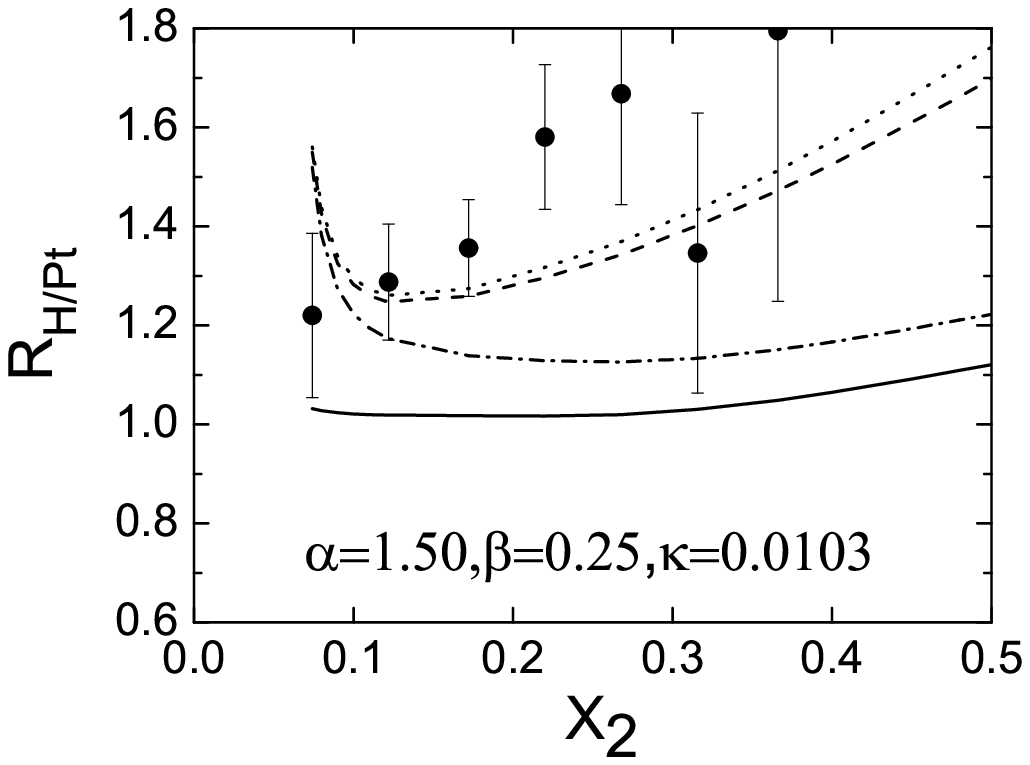} \hspace{0.5mm}
        \includegraphics*[width=70mm]{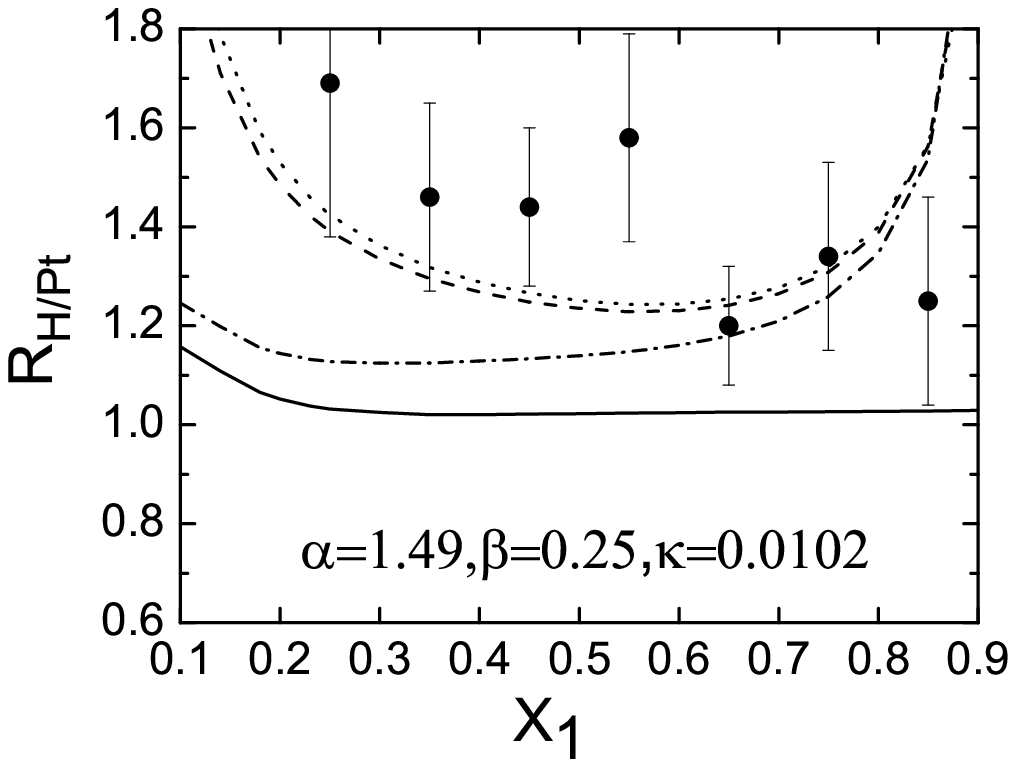} \\
\vspace{-1.0cm} \caption{ The nuclear Drell-Yan cross section ratios
$R_{A_{1}/A_{2}}(x_{1})$ and $R_{A_{1}/A_{2}}(x_{2})$ by using EPS09
nuclear parton distributions.  The other comments are the same as
those in Fig.2.}
\end{figure}

In order to quantify the sensitivity of our computed results on
quark energy loss with respect to the strength of nuclear
corrections to the parton distributions, the global fit analysis is
carried out by using HKN07, nDS and EPS09 parameterizations. The
results given from the three parametrizations of quark energy loss
are summarized in Table IV with the values of $\alpha$, $\beta$ and
$\kappa$,  their corresponding rescaled error, $\chi^2/ndf$ and S
factors. It can be found that the obtained parameter values in three
expressions for quark energy loss are smaller  than that by HKM
nuclear effects. It directly reflects the deviation between HKM
nuclear corrects to sea quark distribution and other three sets. The
calculated results are compared with NA3$^{[14]}$ and NA10a$^{[15]}$
data  including quark energy loss effect and nuclear effects of
parton distributions from HKN07, nDS and EPS09 sets in Fig.3, Fig.4
and Fig.5, respectively.  It is shown from these figures that  the
computed results from the linear quark energy loss are identical to
those from the  quadratic quark energy loss. The theoretical
prediction on $R_{W/D}$ from the incident-parton momentum fraction
quark energy loss exists a significantly large deviation from that
by the linear (or quadratic) quark energy loss in the region $x_2
>0.3$ and $x_1< 0.4$.  The tendency
of $R_{W/D}$ as a function of quark momentum fraction does not
support the possibility of the incident-parton momentum fraction
quark energy loss.   Additionally, it can be found that with
independence on the nuclear modification of parton distributions,
NA3 experiment rules apparently out the incident-parton momentum
fraction quark energy loss. In view of the large experimental error
in NA3 and NA10 data,  it is desirable to operate precise
measurements on the nuclear Drell-Yan reactions from lower incident
beam energy.

\section*{5 Summary and concluding remarks}

In summary, the available data on nuclear Drell-Yan differential
cross section ratio as a function of the quark momentum fraction
have been analysed with three parametrizations of quark energy loss
and four typical sets of nuclear parton distribution functions. It
is found that with independence on the nuclear modifications of
parton distributions, the experimental data from lower incident beam
energy rule out the possibility of the incident-parton momentum
fraction quark energy loss. The existing experimental data do not
distinguish between the linear and quadratic dependence of quark
energy loss. It is worth to mention that the mean energy loss is
employed in our calculations. In hot and dense matter, however, the
mean energy loss of the highly energetic partons would be considered
very simplistic. Rather, it is now accepted that at least  the
probability distribution $P(\Delta E, L)$ of energy loss $\Delta E$
given a path $L$  is the relevant quantity, which then needs to be
averaged over geometry, i.e. a calculation needs to include
explicitly both dynamical fluctuations given the same path, and
fluctuations of the path a quark takes through the medium. It is
possible that the distinction between linear and quadratic energy
loss as observed in present analysis is lost by not accounting for
fluctuations. However, while the quadratic dependence of energy loss
is argued to arise from the Landau-Pomeranchuk-Migdal effect, it is
now known that effectively due to finite energy corrections even an
Landau-Pomeranchuk-Migdal-driven radiative energy loss reverts to an
approximately linear dependence quickly$^{[32, 33]}$.

From the global fit of all selected data, we obtain the quark energy
loss per unit path length $\alpha =1.21 \pm 0.09 $ GeV/fm  by  HKM
nuclear parton distribution functions. By combining our previous
discussion on the semi-inclusive deep inelastic scattering of lepton
on nuclear targets$^{[10]}$, our result on the mean energy loss per
unit length of an incoming quark is not in support of the
theoretical prediction: the mean energy loss of an outgoing quark is
three times larger than that of an incoming quark approaching the
medium$^{[34]}$. In ultra-relativistic heavy-ion collisions,
however, the medium-modified fragmentation function is obtained from
a computation of an in-medium parton-shower followed by
hadronization$^{[35, 36]}$, i.e. what matters for the final state is
not the energy loss of a single quark but rather the modified
development of a parton shower. This virtuality evolution may
explain the difference between the energy loss of incoming and
outgoing quarks - while incoming quarks all in all probably can be
considered on-shell, outgoing quarks are significantly off-shell due
to the hard scattering.

In addition, the obtained value of the parameter in the quark energy
loss expression from HKM nuclear effects is larger than that using
other three sets of nuclear  parton distribution.   Our calaulated
results show that the energy loss effect of the incident quark has a
distinct impact on the Drell-Yan cross section. It directly brings
about an overestimation for nuclear correct to the sea quark
distribution if leaving the quark energy loss effect out. Besides,
the Drell-Yan single differential cross section as a function of the
target parton momentum fraction is dominated by nuclear sea and
valance quark distribution, which is similar to the nuclear
structure function in charged-lepton deep inelastic scattering on
nuclei. In order to make the flavor decomposition  of nuclear parton
distribution functions, we need to resort to the neutrino deep
inelastic scattering data. Several works have studied the nuclear
effects in the neutrino-nucleus charged-current inelastic scattering
process$^{[37,38,39]}$.  We suggest  that the new global analysis of
nuclear parton distribution functions should employ the available
experimental data on structure function from neutrino and
charged-lepton deep inelastic scattering on nuclei.

It is worth noting that the fractional energy loss does not provide
a good description of the data in hot-dense matter physics$^{[40]}$
because the  fast parton propagation in cold nuclear and hot-dense
matter contains different physics$^{[41, 42]}$. Moreover, in our
present work, the used experimental data from NA10 Collaboration
recorded the $x_2$ dependence of nuclear Drell-Yan  cross section
ratio from $ 0.12 < x_2 < 0.45$, which can be covered by Fermilab
E906/SeaQuest experiment$^{[43]}$. Therefore, we desire that  our
paper can provide useful reference for E906's insight on the enengy
loss of an incoming quark propagating in cold nucleus.

\vskip 0.2cm
{\bf Acknowledgments}
This work was supported in part by the National Natural Science
Foundation of China(11075044) and  Natural Science Foundation of
Hebei Province(A2008000137).

\end{document}